\documentclass[12pt,preprint]{aastex62}
\usepackage{color}
\usepackage{graphicx}
\usepackage{longtable}
\usepackage{multirow}
\usepackage{url}
\usepackage{subfigure}
\usepackage{amsmath}
\usepackage{lipsum}

\begin{document}
	
\title{Understanding FRB 200428 in the synchrotron maser shock model: consistency and possible challenge}

	\author[0000-0002-7555-0790]{Q. Wu}
	\affiliation{School of Astronomy and Space Science, Nanjing University, Nanjing 210093, China}
	
	\author[0000-0001-6545-4802]{G. Q. Zhang}
	\affiliation{School of Astronomy and Space Science, Nanjing University, Nanjing 210093, China}
	
    \author[0000-0003-4157-7714]{F. Y. Wang} \affiliation{School of Astronomy and Space Science, Nanjing
	University, Nanjing 210093, China} \affiliation{Key Laboratory of
	Modern Astronomy and Astrophysics (Nanjing University), Ministry of
	Education, Nanjing 210093, China}

	\author{Z. G. Dai}
	\affiliation{School of Astronomy and Space Science, Nanjing
		University, Nanjing 210093, China} \affiliation{Key Laboratory of
		Modern Astronomy and Astrophysics (Nanjing University), Ministry of
		Education, Nanjing 210093, China}

	\correspondingauthor{F. Y. Wang}
	\email{fayinwang@nju.edu.cn}
	
\begin{abstract}
Recently, the discovery of Galactic FRB 200428 associated with a X-ray burst (XRB) of SGR 1935+2154
has built a bridge between FRBs and magnetar activities. In this paper, we assume that the XRB occurs in the magnetar magnetosphere. We show that the observational properties of
FRB 200428 and the associated XRB are consistent with the predictions of synchrotron maser emission at ultrarelativistic
magnetized shocks, including radiation efficiency, similar energy occurrence frequency distributions, and simultaneous arrive times. It requires that the upstream medium is a mildly relativistic baryonic
shell ejected by a previous flare. The energy injection by flares responsible for the radio bursts will produce a magnetar wind nebula, which
has been used to explain the persistent radio source associated FRB 121102.
We find that the radio continuum around SGR 1935+2154 can be well understood in the magnetar wind nebula model, by assuming the same
energy injection rate $\dot{E} \propto t^{-1.37}$ as FRB 121102. The required baryonic mass is also estimated form the observations of FRB 121102 by GBT and FAST.
By assuming the same radiation efficiency $\eta \sim 10^{-5}$, the total baryonic mass ejected from the central magnetar is about 0.005 solar mass. This value is much larger than the typical mass of a magnetar outer crust, but is comparable to the total mass of a magnetar crust. 
\end{abstract}


\section{Introduction}
Fast radio bursts (FRBs) are short and intense burst of radio waves that suddenly appear in the distant universe,
with a few milliseconds duration \citep{Lorimer2007,Thornton2013}. Most of them are extragalactic
events because their large dispersion measures (DM) far exceed the contribution of the Milky Way \citep{Petroff2019,Cordes2019}.
Although nine FRBs have been localized, the origin of FRB are still unknown. Many theoretical models of FRBs have been
proposed \citep{Platts2019}.  The millisecond durations and huge energy releases of FRBs are suggestive of their
central engines being stellar mass compact objects such as magnetars \citep{Popov2013,Kulkarni2014,Murase2016,Katz2016,Metzger2017,Wang2017,Beloborodov2017,Lu2018,Yang2018,Wadiasingh2019,Wang2020}.
The high linear polarization and large rotation measures of FRB 121102 also require a strongly-magnetized central engine
and environment \citep{Michilli2018}. Meanwhile, the statistical similarity, such as energy, duration and waiting time, between FRB 121102 and Galactic magnetar
flares, supports the magnetar central engine \citep{Wang2017,Wadiasingh2019,Cheng2020}.

On 28 April 2020,  FRB 200428 was observed to be emitted from the Galactic magnetar
SGR 1935+2154 \citep{CHIME/FRB2020,Bochenek2020}. The double-peaked burst was temporally coincident with
a double-peaked XRB with the same time separation \citep{Li2020,Ridnaia2020}. This association confirms the connection between FRBs and flare activities of magnetars.
Interestingly, it is well known that solar type III radio bursts are usually associated with solar XRBs \citep{Bastian1998,Reid2014}.
The peak time of XRBs precedes the peaks of solar radio burst by a few seconds \citep{Bastian1998}.
Similar behaviors between FRBs and solar type III radio bursts have been found \citep{ZhangG2019}.
There are some theoretical works on this FRB \citep{Margalit2020,Lu2020,Lyutikov2020,Dai2020,Geng2020,Katz2020,Yu2020,Yang2020}.
In this paper, we focus on the synchrotron maser shock model proposed by \citet{Metzger2019}. It has been shown that
this model can explain the properties of cosmological FRBs \citep{Metzger2019,Margalit2020a}. Whether FRB 200428 can be explained by this model
is still under debate \citep{Margalit2020,Lu2020,Yu2020,WangJ2020}.

In this letter, we use the latest observations to test the parameters of the synchrotron maser shock model
and discuss its consistency and the challenges. This paper is organized as follows. In section \ref{sec2}, we introduce the observations of FRB 200428 and SGR
1935+2154. In section \ref{sec3}, the parameter constraints on the synchrotron maser model are given.
The energy frequency distributions of FRBs and XRBs of SGR 1935+2154 are discussed in section \ref{sec4}.
In section \ref{sec5}, we discuss the rotation measure and the persistent radio source of FRB 121102 and FRB 200428.
In section \ref{sec6}, the baryon mass required by the synchrotron maser model is estimated.
Summary will be given in section \ref{sec7}.

\section{Observational properties of SGR 1935+2154 and FRB 200428}\label{sec2}

\subsection{SGR 1935+2154}\label{sgr1935}

SGR 1935+2154 is a Galactic magnetar, which sprayed a short
burst first detected by Swift/Burst Alert Telescope (BAT), on 2014 July 5
\citep{Stamatikos2014, Lien2014}. After that, different telescopes, including
Swift/X-ray Telescope, Chandra and XMM-Newton
X-ray observations, carried out continuous observation of X-ray pulsations,
and confirmed that the source is a typical magnetar with a spin period $P \sim 3.24 \ \rm s$,
spin-down rate $\dot{P}  \simeq 1.43\times 10^{-11} \rm{\ s \  s^{-1}}$ and a surface dipolar magnetic
field strength of $B \sim 2.2 \times 10^{14} \ \rm{G} $ \citep{Israel2016}.
This source may be associated with the supernova remnant (SNR) G57.2+0.8.
The distance of SGR 1935+2154 has a large range, i.e., from 4.5 to 12.5 kpc \citep{Israel2016,Kothes2018,Zhou2020,Zhong2020}. In this paper, we adopt the distance of $d = \rm 9 \ kpc$.

Since April 2020, multiple short and bright bursts have made up a “burst forest” which means
SGR 1935+2154 has entered a new active period \citep{Hurley2020,Veres2020}.
Luminous hard X-ray bursts were observed by INTEGRAL \citep{Mereghetti2020},
AGILE \citep{Tavani2020}, Konus-Wind \citep{Ridnaia2020}
and \textit{Insight}-HXMT \citep{Li2020}, respectively.
The light curve of the XRB shows two narrow peaks with an interval of about
30 ms \citep{Ridnaia2020,Li2020}, which is consistent of the separation time between
the two narrow peaks in FRB 200428.
The total fluence of the XRB, measured by Konus-\textit{Wind} in the 20–500 keV band, is
$ F_{X} = (9.7 \pm 1.1) \times  10^{-7}  \rm  erg / cm^{2} $. The peak energy flux
is $ S_X = (7.5 \pm 1.0) \times 10^{-6} \rm \ erg / cm^2 / s$, in a 16 ms time
interval \citep{Ridnaia2020}.
For the distance of $d = \rm 9 \ kpc$, the burst energy in
X-rays is
\begin{equation}\label{eq_E_flare}
E_{\rm X} = 4 \pi F_{X} d^2 = 9.4 \times 10^{39} \ \rm erg .
\end{equation}
The burst spectrum can be fitted with a black-body plus power-law model \citep{Li2020}.

\subsection{FRB 200428}\label{FRB200428}

On April 28, 2020, \cite{CHIME/FRB2020} detected a luminous millisecond radio
burst at 400-880 MHz, which was considered to be spatially and temporally coincident with the hard XRB from SGR 1935+2154 \citep{CHIME/FRB2020}.  Meanwhile, the Survey for Transient Astronomical Radio Emission 2 (STARE2) radio array also observed
this millisecond-duration radio burst \citep{Bochenek2020}.
The radio burst detected by the Canadian Hydrogen Intensity Mapping Experiment (CHIME) radio
telescope at 400-800 MHz,  consists of two sub-bursts with best-fit temporal widths of $0.585 \pm 0.014$ ms
and $0.335 \pm 0.007$ ms separated by $28.91 \pm 0.02$ ms \citep{CHIME/FRB2020}.
According to \cite{CHIME/FRB2020}, the average fluence of 480 kJy ms for the first
component and 220 kJy ms for the second were reported. The band-average peak flux density was 110 kJy
for the first component and 150 kJy for the second, respectively.
The value of DM along the direction of FRB 200428 is $ 332.7206 \pm 0.0009 \rm \ pc \ cm^{-3}$.

Another observation of FRB 200428 comes from the STARE2
radio array in the 1281–1468 MHz band \citep{Bochenek2020}.
The burst was detected with an intrinsic width of 0.61 ms. The band-averaged fluence is
$1.5 \times 10^6$ Jy ms.
The isotropic energy of FRB 200428 observed by STARE2 is
\begin{equation}\label{E_FRB}
E_{\rm FRB} = 4\pi F_{\nu}d^2\nu_{\rm c}=2\times 10^{35} \rm erg,
\end{equation}
where $F_{\nu}$ is the fluence of FRB 200428 and the center frequency of STARE2 $\nu_{\rm c} = 1.4 \  \rm GHz$ \citep{Bochenek2020}.
The isotropic-equivalent energies of two sub-bursts detected by CHIME are $1.9 \times 10^{34}$ erg and
$8.5 \times 10^{33}$ erg, respectively. Here we assume the instrument bandwidth of CHIME is 400 MHz.
The brightness temperature $10^{32}$ K is similar to that of cosmological
FRBs \citep{Bochenek2020}. It has been confirmed that FRB 200428 is temporally and spatially consistent with XRB and they both have the sub-structure \citep{Li2020,Ridnaia2020}.
The energy ratio between FRB 200428 and the XRB from SGR 1935+2154 is
\begin{equation}\label{eq_eta}
\eta \equiv \frac{E_{\rm FRB}}{E_{\rm X}} \sim 10^{-5}.
\end{equation}

\section{Constraints on model parameters}\label{sec3}
For the synchrotron  maser shock model, the radiation can be powered by a relativistic shock propagating into a moderately magnetized ($\sigma>10^{-3}$) upstream medium \citep{Lyubarsky2014,Beloborodov2017,Metzger2019}.
Nevertheless, the properties of shocks and the composition of the upstream medium are widely discussed.
So far, three cases have been proposed, including the upstream is a rotationally-powered pulsar wind
\citep{Beloborodov2017}, a magnetar wind nebula \citep{Lyubarsky2014}, or a baryonic shell \citep{Metzger2019}.
In the model of \citet{Metzger2019}, during a magnetar XRB, the ejecta with two components will be launched.
The initial ultrarelativistic ($\Gamma_{ej}$) one with high magnetization may be driven by the engine that triggers the XRB to power the FRB.
This ultrarelativistic component may consists of electron-positron pairs, as suggested by \citet{Yu2020}. The other is a more prolonged phase of ion-loaded mass-loss with a sub-relativistic velocity $\beta_w=v_w/c<1$, which provides the upstream medium to generate the FRB.

We use the observations of FRB 200428 and the associated XRB to constrain
the parameters of this model. Form the observations, the XRB associated with FRB
200428 is similar to other XRBs of SGR 1935+2154 \citep{Lin2020}. Therefore, we assume that
it occurs in the magnetosphere to trigger the mass ejection. Unlike \citet{Margalit2020}, they assumed
that the XRB is mainly produced by the relativistic-hot electrons, similar as gamma-ray burst afterglows. The radiation efficiency $\zeta$ for converting the kinetic energy of the ejecta into
coherent radio radiation in the baryonic shell can be simulated as \citep{Metzger2019}
\begin{align}\label{eq_zeta}
\zeta \approx  4.8 \times 10^{-4}
\left(\frac{E_{\rm X}}{9.4 \times 10^{39} \rm \ erg}\right)^{-4/5}
\left(\frac{E_{\rm FRB}}{2 \times 10^{35} \rm \ erg}\right)^{4/5}
\left(\frac{m_{*}}{m_{\rm e}}\right)^{1/2}
\left(\frac{f_{\rm e}}{0.5}\right)^{-1/4}   \nonumber \\
\times
\left(\frac{\nu_{\rm FRB}}{1.4 \rm \ GHz}\right)^{1/4}
\left(\frac{ t_{\rm FRB}}{0.61\rm \ ms}\right)^{1/4},
\end{align}
where $E_{\rm X}$ and $E_{\rm FRB}$ are the energy releases of XRB and FRB 200428 detected by STARE2,
respectively. In the following, the observation of STARE2 is used.
We assumed that the energy of XRB approximate equals to the energy carried by the ultrarelativistic ejecta.
$m_*$ is the effective mass and $f_{\rm e}$ is defined as the ratio of electron density to external particle
density, i.e.  $f_{\rm e} \equiv n_{\rm e} / n_{\rm ext}$. For the upstream medium of electron-ion composition,
we suppose $f_{\rm e} \simeq 0.5$.
The predicted radiation efficiency $\zeta$ can explian the observed one (equation \ref{eq_eta}).

We assume the instrict synchrotron maser efficiency  $f_{\xi} = 10^{-3} $ \citep{Metzger2019}. Using the observed frequency $\nu_{\rm FRB}$, fluence $F_{\nu}$ and duration $t_{\rm FRB}$, we can solve the Lorentz factor $\Gamma$ of the shocked gas, the radius of shock $ r_{\rm sh}$ and the external density of the medium at $r_{\rm sh}$ as follows
\begin{align}\label{eq_gamma}
\Gamma \approx  53.4
\left(\frac{m_*}{m_e}\right)^{1/30}
\left(\frac{f_e}{0.5}\right)^{1/15}
\left(\frac{f_{\xi}}{10^{-3}}\right)^{-1/15}
\left(\frac{\nu_{\rm FRB}}{1.4 \rm\ GHz}\right)^{-7/30}    \nonumber \\
\times
\left(\frac{ t_{\rm FRB}}{0.61\rm\ ms}\right)^{-2/5}
\left(\frac{E_{\rm FRB}}{2 \times 10^{35} \rm \ erg}\right)^{1/6},
\end{align}

\begin{align}\label{eq_r}
r_{sh}\simeq 2\Gamma^2ct \approx  9.9 \times 10^{10}\  \mathrm{cm}
\left(\frac{m_*}{m_e}\right)^{1/15}
\left(\frac{f_e}{0.5}\right)^{2/15}
\left(\frac{f_{\xi}}{10^{-3}}\right)^{-2/15}  \nonumber \\
\times
\left(\frac{\nu_{\rm FRB}}{1.4 \rm \ GHz}\right)^{-7/15}
\left(\frac{ t_{\rm FRB}}{0.61\rm\ ms}\right)^{1/5}
\left(\frac{E_{\rm FRB}}{2 \times 10^{35} \rm \ erg}\right)^{1/3},
\end{align}

\begin{align}\label{eq_n}
n_{\rm ext}(r_{\rm sh})\approx   1.8 \times 10^{4} \ \mathrm{cm^{-3}}
\left(\frac{m_*}{m_e}\right)^{2/15}
\left(\frac{f_e}{0.5}\right)^{-11/15}
\left(\frac{f_{\xi}}{ 10^{-3}}\right)^{-4/15}  \nonumber \\
\times
\left(\frac{\nu_{\rm FRB}}{1.4 \rm\ GHz}\right)^{31/30}
\left(\frac{ t_{\rm FRB}}{0.61\rm\ ms}\right)^{2/5}
\left(\frac{E_{\rm FRB}}{2 \times 10^{35} \rm \ erg}\right)^{-1/3}.
\end{align}

The induced Compton scattering (ICS) will suppress the short radio bursts.  Only the optical depth of
the induced Compton scattering $\tau_{\rm ICS}$ is smaller than 3, radio radiation can escape from the baryonic shell
\citep{Metzger2019}. When the optical depth equals to 3, the corresponding frequency is defined as the peak frequency of
the observed spectrum, i.e $\nu_{\rm max} \equiv \nu(\tau_{\rm ICS}=3)$. Thus, $\nu_{\rm max}$ can be expressed as
\begin{align}\label{eq_v_max}
\nu_{\mathrm{max}} =    0.97  \ \mathrm{GHz}
\left(\frac{f_{\xi}}{10^{-3}}\right)^{1 / 4}
\left(\frac{E_{\rm FRB}}{2 \times 10^{35} \rm \ erg}\right)^{5 / 32}
\dot{M}_{21}^{15 / 32}\left(\frac{\beta_{w}}{0.3}\right)^{-45 / 32}  \nonumber \\
\times
\Delta T_{4}^{-30 / 32}
\left(\frac{ t_{\rm FRB}}{0.61\rm\ ms}\right)^{-7 / 32} (\frac{f_e}{0.5})^{5/8},
\end{align}
where $\Delta T=10^4$ s is the time interval between the successive XRBs and $\beta_{w} = 0.3$ is assumed.

From observations, the separation time of the two narrow X-ray peaks is consistent with that of the two peaks in FRB 200428 \citep{Li2020}.
After corrected the dispersion time delay by 333 pc cm$^{-3}$, the two X-ray peaks each
occur within about $t_\delta \sim$1 ms of the corresponding CHIME burst peaks \citep{CHIME/FRB2020,Li2020}.
Therefore, it requires that the Lorentz factor $\Gamma_{\rm ej}$ of the ultrarelativistic component satisfies

\begin{equation}
\Gamma_{\rm ej} \geq \left(r_{sh}/2c t_\delta \right)^{1/2}\simeq 40.7 \  \left(\frac{r_{\mathrm{sh}}}{9.9\times10^{10} \ \mathrm{cm}}\right)^{1/2} \left(\frac{t_{\delta}}{1\ \rm ms}\right)^{1/2} .
\end{equation}

Incoherent synchrotron afterglow is also produced by the
relativistic-hot electrons, which is similar to the gamma-ray burst afterglow \citep{Lyubarsky2014,Metzger2019}.
The peak frequency of the synchrotron afterglow is given by \citep{Metzger2019}
\begin{align}\label{eq_hv}
h \nu_{\mathrm{syn}} \approx  552.6~\mathrm{keV}
\left(\frac{E_{\rm X}}{9.4\times 10^{39} \rm erg}\right)^{1 / 2}
\left(\frac{t}{1\rm ms}\right)^{-3 / 2}
\left(\frac{\sigma}{10^{-2}}\right)^{1/2},
\end{align}
where $\sigma = 10^{-2}$ and $t$ is the time since the XRB peak.
It's obvious that this peak frequency is similar as that of the XRB occurring in the magnetar's magnetosphere.
Therefore, this emission may overlap the original XRB in the magnetosphere.
\citet{Margalit2020} assumed the XRB is mainly from the synchrotron afterglow. The observational properties of FRB 200428 can be well explained in these two physical scenarios. The radio burst properties are similar in the magnetosphere-powered situation and the shock-powered situation. The predicted shock-powered X-ray fluence is close to the observed one \citep{Margalit2020}, which indicates that part of the XRB may come from shock emission. Thus, the magnetospheric XRBs and the shock-powered XRBs may coexist in this model.

The relative time delay $t_{\delta}$ between the XRB and the radio burst is about $r_{\rm sh}/\Gamma_{\rm ej}^2 c \sim \rm ms$, which is comparable to the radio burst duration. From observations, CHIME \citep{CHIME/FRB2020} and Insight-HXMT \citep{Li2020} found that the two X-ray peaks each occurred within 1 ms of the arrival times of the radio peaks detected by CHIME, which is consistent with our estimation.

\section{Energy frequency distributions}\label{sec4}
In the synchrotron maser model proposed by \citet{Metzger2019}, the energy of FRBs is proportional to the kinetic energy carried by
ultrarelativistic ejetca, if the upstream magnetization is similar. The kinetic energy of ultrarelativistic ejetca has the same order of that of
the corresponding XRB \citep{Yu2020}.
For a magnetar, the magnetization of upstream medium may not change significantly. Therefore, it is expected that the occurrence frequency
distributions of energy for XRBs and the associated FRBs are similar. Only a portion of XRBs is associated with FRBs. For example,
no bursts is found from SGR 1935+2154 by eight-hour observation of FAST \citep{Lin2020}. However, there are 29 XRBs of SGR 1935+2154 in the same period \citep{Lin2020}. The possible reason is the beaming emission of FRBs.

The cumulative energy distribution of FRB 121102 was considered to have a power-law form. \citet{Law2017} found that the slope of  differential distribution of energy is $\sim 1.7$ using the multi-telescope detection. \citet{Gourdji2019} found the cumulative distribution of burst energies with a slope of $1.8 \pm 0.3$ .
Furthermore, \citet{Wadiasingh2019} obtained a differential energy distribution with power law index of $2.3 \pm 0.2$ using 72 bursts.

We compare the energy occurrence distribution between FRBs and XRBs of SGRs. The data of FRB 121102 \citep{ZhangY2018} and magnetar SGR 1935+2154 \citep{Lin2020a} are used. We use a power-law distribution with a high-energy cutoff to fit the cumulative distribution, which reads
\begin{equation}\label{Edis}
N(>E)=A(E^{1-\alpha_E}-E_{\rm max}^{1-\alpha_E}),
\end{equation}
where $\alpha_E$ is the power-law index and $E_{\rm max}$ is the maximum energy of the FRB.
Using equation (\ref{Edis}), \citet{Cheng2020} found that the power-law index of the energy frequency distribution of 93 bursts of FRB 121102 is $\alpha_E=1.63\pm 0.06$. For SGR 1935+2154, the 112 bursts observed by Fermi/GBM from 2014 to 2016 are used.
The power-law index of energy distribution for SGR 1935+2154 is $\alpha_E=1.71\pm 0.03$.
The two distributions are consistent with each other at $1\sigma$ confidence level, which supports the association between XRBs and
FRBs.

\section{Rotation measure and persistent radio source}\label{sec5}
The intermittent injection of ejecta from magnetar will generate an expanding magnetized electron–ion nebula.  Observations show that a luminous ($\nu L_\nu=10^{39}$ erg s$^{-1}$) persistent
radio source coincident to within $\leq40$ pc of the
FRB 121102 location \citep{Marcote2017}. Meanwhile, the high rotation measure (RM) of the bursts, RM$\sim 10^5$ rad m$^{-2}$ is found \citep{Michilli2018}.
The persistent radio emission and high RM may originate from the same medium, showing that the FRB source is embedded
in a dense magnetized plasma \citep{Michilli2018}. The persistent radio source is thought to be synchrochon radiation of the magnetar wind nebula
\citep{Dai2017,Kashiyama2017,Metzger2017}. For the Galactic magnetar SGR 1935+2154, the extended X-ray
emission was found by \citet{Israel2016} and can be interpreted as a pulsar wind nebula.
But \citet{Younes2017} found that there was no extended emission around SGR 1935+2154 using Chandra data. For the XMM-Newton data, they found similar results as \citet{Israel2016}. So there is marginal evidence of the existence of a magnetar wind nebula around SGR 1935+2154. Below, we assume the magnetar wind nebula is around the source.
\citet{Kothes2018} discovered a bright radio shell consists of two narrow arc-like features and a radial magnetic field around SGR 1935+2154, which could be explained by a pulsar wind nebula \citep{Kothes2018}. On the other hand, it could also be explained as interaction of the SNR with
the ambient medium. Here, we only consider the former case. The spectrum of the radio emission is $S_\nu \propto \nu^{-0.55\pm 0.02}$ \citep{Kothes2018}, which is similar to that of the persistent radio source of FRB 121102 ($\nu^{-0.27\pm 0.24}$) \citep{Marcote2017}. The luminosity of the radio shell associated with SGR 1935+2154 is $5\times 10^{33}$ erg~s$^{-1}$ \citep{Kothes2018}. The RM toward the direction of SGR 1935+2154 is about 116$\pm 2\pm 5$ rad m$^{-2}$ \citep{CHIME/FRB2020,FAST2020}.

\begin{figure*}\label{fig1}
	\centering
	\includegraphics[width=\linewidth]{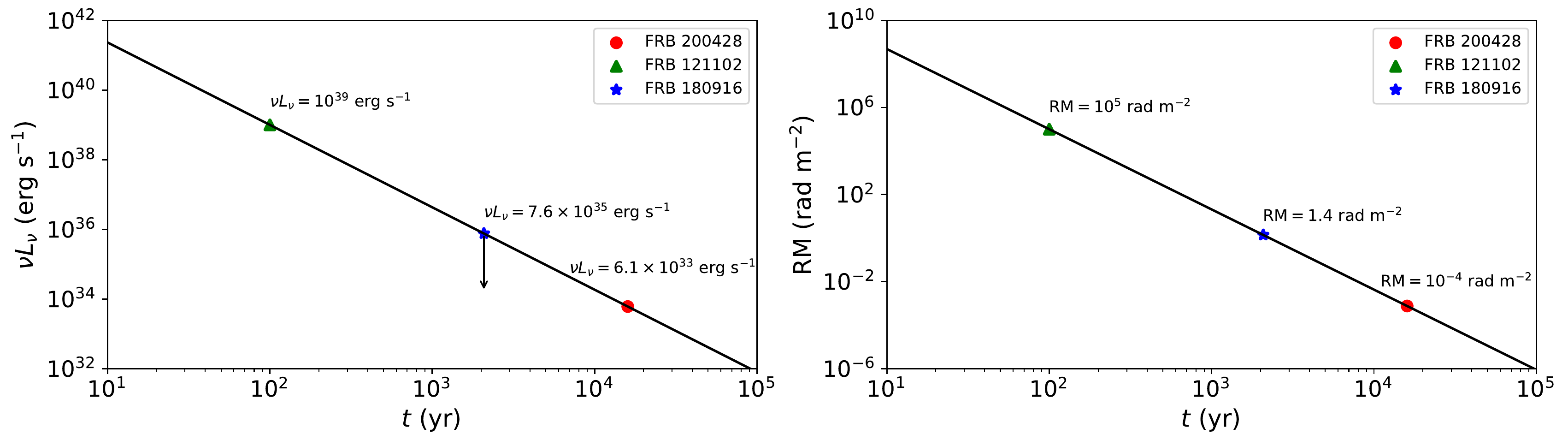}
	\caption{Time evolution of persistent radio source luminosity ($\nu L_{\nu}$) (left panel) and RM value (right panel). Red dot for FRB 200428, green triangle for FRB 121102 and blue pentagon for FRB 180916.
	In the left panel, the blue pentagon represents the upper limit of the persistent radio source luminosity of FRB 180916 from the VLA data. In order to explain the upper limit of the persistent radio source luminosity of FRB 180916 from equation \ref{vLv}, the central magnetar age must be larger than about 2000 yr. Using this lower limit of central magetar age for FRB 180916, the model-predicted RM (equation \ref{RM}) is about 1.4 $\rm rad \ m^{-2}$, which is consistent with observation.}
	\label{vL/RM}
\end{figure*}

It has been shown that the persistent radio source associated with FRB 121102 can be explained by a single expanding magnetized electron–ion wind nebula created by magnetar flares \citep{Margalit2018}.
Assuming that the radio shell emission around SGR 1935+2154 is also powered by pulsar wind nebula, we explain the observations (radio luminosity and RM) using the model proposed by \citet{Margalit2018}.
Considering the magnetar releases its free magnetic energy into the nebula in a power-law form \citep{Margalit2018}
\begin{equation}
\dot{E}\propto t^{-\alpha},
\end{equation}
the time evolution of RM is
\begin{equation}\label{RM}
\rm RM \propto t^{-(6+\alpha)/2},
\end{equation}
and the decay of radio source luminosity is
\begin{equation}\label{vLv}
\nu L_\nu \propto t^{-(\alpha^2+7\alpha-2)/4}.
\end{equation}
The age of the magnetar powering FRB 121102 is found to be very young, i.e., a few decades to 100 years \citep{Metzger2017,Cao2017,Kashiyama2017}.
We take 100 years as a fiducial value. So the luminosity of the magnetar wind nebula is $10^{39}$ erg s$^{-1}$ at $t=100$ yr. The age of SGR 1935+2154 is about 16,000 years from non-detection of thermal X-ray emission from the supernova remnant and the relatively dense environment \citep{Zhou2020}.
In order to explain the observations of RM and radio source luminosity for FRB 121102, $\alpha=1.3-1.8$ is adopted \citep{Margalit2018}.
For $\alpha=1.37$, the persistent radio source luminosity is about $\nu L_\nu=6.1\times 10^{33}$ erg s$^{-1}$ at $t=16,000$ years, which is
dramatically consistent with the observed one of SGR 1935+2154. For RM, the value is 8$\times 10^{-4}$ rad m$^{-2}$ at $t=16,000$ years.
It is also consistent with the FAST observation. From a highly polarized radio burst from SGR 1935+2154, it is found that the RM contribution from the local magnetoionic environment is very low \citep{FAST2020}.
In figure \ref{vL/RM}, we show the time evolution of the persistent radio source luminosity and RM for FRB 121102 and FRB 200428.

Figure \ref{vL/RM} also show the observations of FRB 180916 as blue pentagons, which was located in a nearby massive spiral galaxy \citep{Marcote2020}.
From the VLA data, the persistent radio source luminosity of FRB 180916.J0158+65 was constrained to
$\nu L_{\nu} < 7.6 \times 10^{35} \rm \ erg \ s^{-1}$.
Based on the time evolution of the persistent radio source, we find that the lower limit of age for central magnetar is about 2079 yr.
Using this age, the RM is about 1.4 $\rm rad \ m^{-2}$, which is also consistent observations \cite{CHIME/FRB2019}.
\citet{CHIME/FRB2019} measured the RM of FRB 180916 is $-114.6\pm 0.6 \rm \ rad \ m^{-2}$.
The observed $\rm RM_{tot}$ is consists of the contribution of the Milky Way, the host galaxy and the source.
In the direction of the FRB 180916, the contribution of RM from Milky Way is $-115\pm 12  \rm \ rad \ m^{-2}$ \citep{Ordog2019}. Therefore, the major contribution to the RM value of FRB 180916 comes from the Milky Way, while the contribution of the FRB source and the host galaxy can be as small as a few $\rm rad \ m^{-2}$.

Therefore, the RM and radio source luminosity of FRB 121102, FRB 200428 and FRB 180916 can be well understood in a single expanding magnetized electron–ion wind nebula embedded within a supernova remnant.

\section{Baryon mass budget}\label{sec6}
In the model proposed by \citet{Metzger2019}, the FRBs are produced in the collision between the ultra-relativistic
ejecta and the external baryonic shell ejected by the previous flare. The baryonic mass required for producing
one cosmological FRB is about $\Delta M=\dot{M}\Delta T$, where $\dot{M}\sim 10^{19}-10^{21}$ g s$^{-1}$ and $\Delta T\sim 10^4$ s is the time interval between the successive XRBs \citep{Metzger2019}.
For Galactic FRB 200428,  a low value $\Delta M\sim 10^{20}$ g is required \citep{Yu2020}. An important question is whether the magnetar can supply
so much baryonic material. Using the observation of FRB 121102, we give an estimation of the baryonic mass required by this model.

We use the data of FRB 121102 observed by Green Bank Telescope (GBT) \citep{ZhangY2018}, which is the largest sample in a single observation.
There are 93 bursts in 5 hours observation. The energy of each burst can be derived from
\begin{equation}
\label{eq:Energy}
E = 4 \pi d_l^2 F \nu_{\rm c},
\end{equation}
where $d_l$ is the luminosity distance at $z=0.19$, $F$ is the fluence and $\nu_{\rm c}$ is the center frequency of GBT \citep{ZhangY2018}. The $\Lambda$CDM
model with $H_0$ = 67.74 km/s/Mpc, $\Omega_M$ = 0.31 and $\Omega_\Lambda$= 0.69 is used. The total energy release in these
five hours is $E_t = \sum_i E_i$, where $E_i$ the energy of $i$th burst. \citet{Rajwade2020} found a possible
period of 157 days with a duty cycle of 56 percent. If this observation is in the active phase of FRB 121102, the energy release in a period can be derived as
\begin{equation}
\label{eq:ErPeriod}
E_{\rm radio} = \frac{E_t}{T_{\rm obs}}T_{\rm period}\xi\simeq 1.8 \times 10^{43} \rm erg,
\end{equation}
where $T_{\rm obs} = 5 $ hr is the observation time, $T_{\rm period} = 157$ days is the period, and $\xi = 0.56$ is the duty cycle.

\begin{figure*}\label{fig2}
	\centering
	\includegraphics[width=\linewidth]{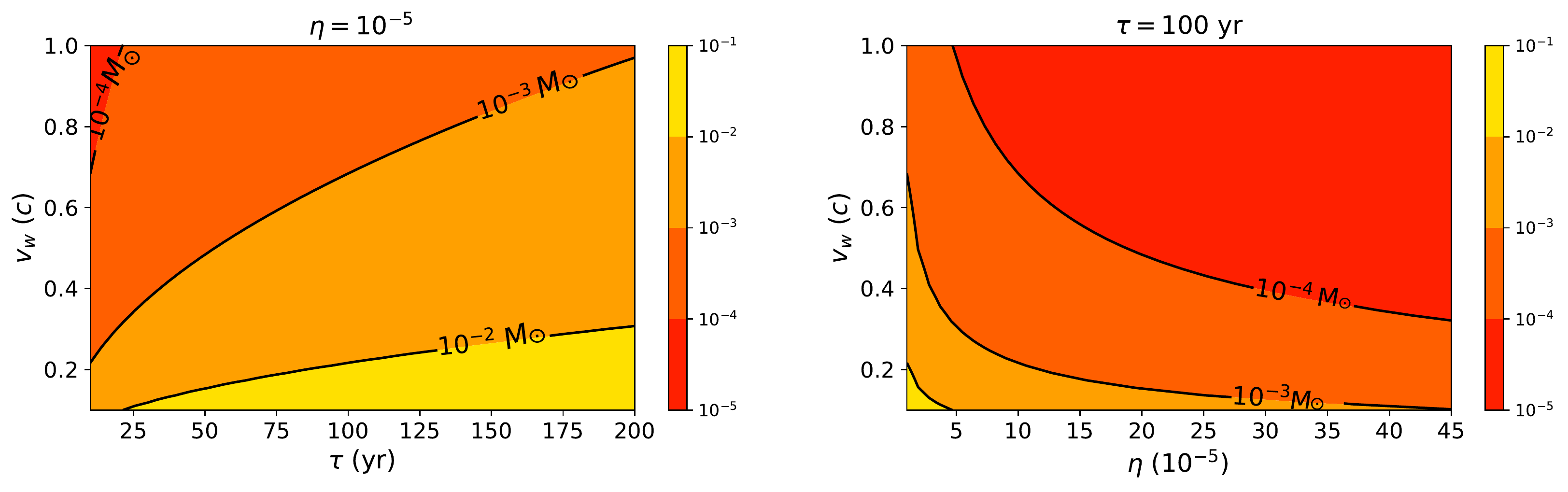}
	\caption{Constraints on the required baryonic mass $M_B$ in the $\tau-v_w$ plane (left panel) and $\eta-v_w$ plane (right panel). In the left panel, $\eta=10^{-5}$ is fixed.
	The value of $M_B$ is larger than $10^{-5}M_\odot$ for the whole range of parameters. In the right panel, $\tau=100$ yr is fixed. The value of $M_B$ is larger than $10^{-5}M_\odot$ for the whole range of parameters. }
	\label{Mb}
\end{figure*}

The observation of FRB 200428 suggests the FRBs accompanied by XRBs. The energy ratio between radio burst and XRB is about
$\eta \sim 10^{-5}$. If this value is valid for FRB 121102, the X-ray energy release in one period for FRB 121102 is about
\begin{equation}
\label{eq:EXReriod}
E_{\rm X} = \frac{E_{\rm radio}}{\eta} \simeq 1.8 \times 10^{48} \ {\rm erg}\ (\frac{\eta}{10^{-5}})^{-1} .
\end{equation}
The typical active timescale of magnetars is about 100 years \citep{Beloborodov16}. The total energy release of XRBs in the active phase is
\begin{equation}
\label{eq:Etot}
E_{\rm total} = \frac{\tau}{T_{\rm period}}E_{\rm X} \simeq  4.2\times 10^{50} \ \mathrm{erg}\ (\frac{\eta}{10^{-5}})^{-1}  (\frac{\tau}{100 \rm yr}) \rm ,
\end{equation}
where $\tau$ is the active timescale of magnetar. For FRB 121102, $\tau\sim 100$ yr is assumed.
Compared to the rotational energy, the magnetic energy is the main reservoir responsible for powering FRBs. The magnetic energy of magnetar is
\begin{equation}
E_B\simeq B^2R^3/6 \approx 3\times 10^{49} ~\rm erg \ B^2_{16} ,
\end{equation}
where $B=10^{16}$ G is the interior magnetic field
strength, and $R=$12 km is the magnetar radius. This value is smaller than the required XRB energy.
The XRB energy is shared by the ultra-relativistic ejecta and the baryon shell with a
sub-relativistic velocity $v_w$.
It has been found that the kinetic energy of baryon shell is comparable to the flare energy \citep{Metzger2019,Margalit2020a}. We assume that the baryon shell is sub-relativistic
with a typical velocity $v_w=0.3c$. Based on above assumptions, the baryonic mass ejected by magnetar in the active time can be derived as
\begin{equation}
\label{eq:massloss}
M_B \simeq 2\frac{E_{\rm total}}{v_{w}^2} \simeq 5.2 \times 10^{-3} \, {M_\odot} \ (\frac{\eta}{10^{-5}})^{-1}  (\frac{\tau}{100 \rm yr}) (\frac{v_{w}}{0.3 c})^{-2}.
\end{equation}
This value is much larger than the typical mass ($10^{-5}\, \mathrm{M_\odot}$) of a magnetar outer crust \citep{Gudmundsson1983,Glendenning1992}. According to the structure of magnetars, 
the core is mainly composed of superfluid. The ejected baryon matter of the magnetar is mainly provided by crust. 
The baryonic mass estimated in this model is larger than the typical mass of a magnetar outer crust, indicating that the outer crust of the magnetar can not eject enough baryonic matter required by the model.  	
In Figure \ref{Mb}, we show the constraints on $M_B$ using equation (\ref{eq:massloss}).
If the radiation efficiency $\eta=10^{-5}$ from FRB 200428 is used, the required baryonic mass $M_B$ is always larger than $10^{-5}~M_\odot$ for the whole parameter ranges as shown in the left panel of Figure \ref{Mb}.
If $\tau=100$ yr is fixed, the same result is shown in the right panel of Figure \ref{Mb}.
The outer crust of magnetars is not sufficient to provide such a large baryonic mass, which challenges the synchrotron maser shock model. Interestingly, if the inner crust is included, the total mass of crust 
is about 0.01 $\mathrm{M_\odot}$ \citep{Chamel2008}, which is larger than the required baryonic mass. Theoretically, the physical mechanism and the rate of the baryonic mass ejection are both uncertain. 
More investigations are required.
We also use the bursts of FRB 121102 observed by FAST to estimate the required baryonic mass \citep{Li2019}. The observation of FAST provides the information of lowest-energy bursts of FRB 121102 so far. In a 56.5 hours observation, 1121 bursts were observed. The total energy of these bursts is 3.14 $ \times 10^{41} $ erg\footnote{D. Li and P. Wang private communications.}. Therefore, the average energy release in one period is about $ 1.18\times 10^{43} $ erg. Using the same formulae as above, the baryon mass can be approximated as
\begin{equation}
	\label{eq:masslossFAST}
	M_B \sim 3.5 \times 10^{-3} \, {M_\odot} \ (\frac{\eta}{10^{-5}})^{-1}  (\frac{\tau}{100 \rm yr}) (\frac{v_{w}}{0.3 c})^{-2},
\end{equation}
which is similar to that derived from GBT observation.

The required baryonic mass can also be roughly derived by $\dot{M}\times \tau$. For $\dot{M}\sim 10^{19}-10^{21}$ g s$^{-1}$ and $\tau=100$ yr, the mass is between $1.5\times 10^{-5}$ to $1.5\times 10^{-3}~M_\odot$, which covers two orders of magnitude. The largest value is comparable to our result, i.e., equation (\ref{eq:massloss}). The smallest value is similar to the mass of magnetar outer crust. However, the bursts of FRB121102 occur in an irregular fashion, and appear to be clustered \citep{Wang2017,Oppermann2018}. So this calculation is an approximation. In order to obtain a precise value, the period and duty circle of  bursts must be considered.

\section{Summary}\label{sec7}

Motivated by the fact that FRB 200428 is spatially and temporally coincident with a hard XRB from
SGR 1935+2154, we test the synchrotron maser shock model in this paper. Our conclusions can be summarized as follows.

(1) Here we consider the case that the upstream medium is electron-ion material. We find that the radiation efficiency, Lorentz factor of shocked gas, the radius of shock and the density external medium are consistent with the upstream medium is an ion-loaded shell released by a recent burst.

(2) Similar energy occurrence frequency distributions between bursts of FRB 121102 and XRBs of SGR 1935+2154 is found, which supports the association of FRBs and XRBs from magnetars.

(3) In this model, the energy injection by intermittent ejecta from magnetar will generate an expanding magnetized electron–ion nebula.
We show that the radio continuum emission around SGR 1935+2154 can be well understood in the magnetar nebula model, by assuming the same energy injection rate $\dot{E} \propto t^{-1.37}$ as FRB 121102.

(4) The RM value contributed by the nebula is $8 \times 10^{-4} \ \mathrm{rad \ m^{-2}}$.
This small value is consistent with the observation that the main contribution of RM is the interstellar medium between SGR 1935+2154 and us.
Therefore, the whole observational properties of FRB 200428 can be well understood in the synchrotron maser shock model.

(5) However, in order to explain the observations, the upstream medium must be an ion-loaded shell.
To study whether the magnetar can provide enough baryonic matter, we use the observation
of FRB 121102 from GBT and FAST to estimate the required baryonic mass.
The baryon mass ejected by the central magnetar in active lifetime is about $5.2\times 10^{-3} \ \rm M_{\odot}$
using the GBT observation, which is much larger than the typical mass of a magnetar outer crust. FAST observation gives a similar result. So, the large baryonic mass challenges this model. If both the outer and inner crusts are considered, the mass of crust is about 0.01 $\mathrm{M_\odot}$, which is enough for the required baryonic mass.


\section*{acknowledgements}
We thank Y.W. Yu, S.L. Xiong, P. Wang, P, Zhou, D. Li. and X.Y. Li for helpful discussions. We thank
the anonymous referee for valuable comments. This work is supported by the National Natural Science Foundation of China (grants U1831207 and 11833003), and the National Key Research and Development
Program of China (grant 2017YFA0402600).

\bibliographystyle{aasjournal}
\bibliography{ms}

\begin{thebibliography}{}
\expandafter\ifx\csname natexlab\endcsname\relax\def\natexlab#1{#1}\fi
\providecommand{\url}[1]{\href{#1}{#1}}

\bibitem[{{Bastian} {et~al.}(1998){Bastian}, {Benz}, \& {Gary}}]{Bastian1998}
{Bastian}, T.~S., {Benz}, A.~O., \& {Gary}, D.~E. 1998, \araa, 36, 131

\bibitem[{{Beloborodov}(2017)}]{Beloborodov2017}
{Beloborodov}, A.~M. 2017, \apjl, 843, L26

\bibitem[{{Beloborodov} \& {Li}(2016)}]{Beloborodov16}
{Beloborodov}, A.~M., \& {Li}, X. 2016, \apj, 833, 261

\bibitem[{{Bochenek} {et~al.}(2020){Bochenek}, {Ravi}, {Belov}, {Hallinan},
  {Kocz}, {Kulkarni}, \& {McKenna}}]{Bochenek2020}
{Bochenek}, C.~D., {Ravi}, V., {Belov}, K.~V., {et~al.} 2020, arXiv e-prints,
  arXiv:2005.10828

\bibitem[{{Cao} {et~al.}(2017){Cao}, {Yu}, \& {Dai}}]{Cao2017}
{Cao}, X.-F., {Yu}, Y.-W., \& {Dai}, Z.-G. 2017, \apjl, 839, L20

\bibitem[{{Chamel} \& {Haensel}(2008)}]{Chamel2008}
{Chamel}, N., \& {Haensel}, P. 2008, Living Reviews in Relativity, 11, 10

\bibitem[{{Cheng} {et~al.}(2020){Cheng}, {Zhang}, \& {Wang}}]{Cheng2020}
{Cheng}, Y., {Zhang}, G.~Q., \& {Wang}, F.~Y. 2020, \mnras, 491, 1498

\bibitem[{{CHIME/FRB Collaboration} {et~al.}(2019){CHIME/FRB Collaboration},
  {Andersen}, {Bandura}, {Bhardwaj}, {Boubel}, {Boyce}, {Boyle}, {Brar},
  {Cassanelli}, {Chawla}, {Cubranic}, {Deng}, {Dobbs}, {Fandino}, {Fonseca},
  {Gaensler}, {Gilbert}, {Giri}, {Good}, {Halpern}, {Hill}, {Hinshaw},
  {H{\"o}fer}, {Josephy}, {Kaspi}, {Kothes}, {Landecker}, {Lang}, {Li}, {Lin},
  {Masui}, {Mena-Parra}, {Merryfield}, {Mckinven}, {Michilli}, {Milutinovic},
  {Naidu}, {Newburgh}, {Ng}, {Patel}, {Pen}, {Pinsonneault-Marotte}, {Pleunis},
  {Rafiei-Ravandi}, {Rahman}, {Ransom}, {Renard}, {Scholz}, {Siegel}, {Singh},
  {Smith}, {Stairs}, {Tendulkar}, {Tretyakov}, {Vanderlinde}, {Yadav}, \&
  {Zwaniga}}]{CHIME/FRB2019}
{CHIME/FRB Collaboration}, {Andersen}, B.~C., {Bandura}, K., {et~al.} 2019,
  \apjl, 885, L24

\bibitem[{{Cordes} \& {Chatterjee}(2019)}]{Cordes2019}
{Cordes}, J.~M., \& {Chatterjee}, S. 2019, \araa, 57, 417

\bibitem[{{Dai}(2020)}]{Dai2020}
{Dai}, Z.~G. 2020, \apjl, 897, L40

\bibitem[{{Dai} {et~al.}(2017){Dai}, {Wang}, \& {Yu}}]{Dai2017}
{Dai}, Z.~G., {Wang}, J.~S., \& {Yu}, Y.~W. 2017, \apjl, 838, L7

\bibitem[{{Geng} {et~al.}(2020){Geng}, {Li}, {Li}, {Xiong}, {Kuiper}, \&
  {Huang}}]{Geng2020}
{Geng}, J.-J., {Li}, B., {Li}, L.-B., {et~al.} 2020, arXiv e-prints,
  arXiv:2006.04601

\bibitem[{{Glendenning} \& {Weber}(1992)}]{Glendenning1992}
{Glendenning}, N.~K., \& {Weber}, F. 1992, \apj, 400, 647

\bibitem[{{Gourdji} {et~al.}(2019){Gourdji}, {Michilli}, {Spitler}, {Hessels},
  {Seymour}, {Cordes}, \& {Chatterjee}}]{Gourdji2019}
{Gourdji}, K., {Michilli}, D., {Spitler}, L.~G., {et~al.} 2019, \apjl, 877, L19

\bibitem[{{Gudmundsson} {et~al.}(1983){Gudmundsson}, {Pethick}, \&
  {Epstein}}]{Gudmundsson1983}
{Gudmundsson}, E.~H., {Pethick}, C.~J., \& {Epstein}, R.~I. 1983, \apj, 272,
  286

\bibitem[{{Hurley} {et~al.}(2020){Hurley}, {Mitrofanov}, {Golovin}, {Litvak},
  {Sanin}, {Kozlova}, {Golenetskii}, {Aptekar}, {Frederiks}, {Svinkin},
  {Cline}, {Goldstein}, {Briggs}, {Wilson-Hodge}, {von Kienlin}, {Zhang},
  {Rau}, {Savchenko}, {E. Bozzo}, {Ferrigno}, {Barthelmy}, {Cummings}, {Krimm},
  {Palmer}, {Boynton}, {Fellows}, {Harshman}, {Enos}, \& {Starr}}]{Hurley2020}
{Hurley}, K., {Mitrofanov}, I.~G., {Golovin}, D., {et~al.} 2020, GRB
  Coordinates Network, 27625, 1

\bibitem[{{Israel} {et~al.}(2016){Israel}, {Esposito}, {Rea}, {Coti Zelati},
  {Tiengo}, {Campana}, {Mereghetti}, {Rodriguez Castillo}, {G{\"o}tz},
  {Burgay}, {Possenti}, {Zane}, {Turolla}, {Perna}, {Cannizzaro}, \&
  {Pons}}]{Israel2016}
{Israel}, G.~L., {Esposito}, P., {Rea}, N., {et~al.} 2016, \mnras, 457, 3448

\bibitem[{{Kashiyama} \& {Murase}(2017)}]{Kashiyama2017}
{Kashiyama}, K., \& {Murase}, K. 2017, \apjl, 839, L3

\bibitem[{{Katz}(2016)}]{Katz2016}
{Katz}, J.~I. 2016, \apj, 826, 226

\bibitem[{{Katz}(2020)}]{Katz2020}
---. 2020, arXiv e-prints, arXiv:2006.03468

\bibitem[{{Kothes} {et~al.}(2018){Kothes}, {Sun}, {Gaensler}, \&
  {Reich}}]{Kothes2018}
{Kothes}, R., {Sun}, X., {Gaensler}, B., \& {Reich}, W. 2018, \apj, 852, 54

\bibitem[{{Kulkarni} {et~al.}(2014){Kulkarni}, {Ofek}, {Neill}, {Zheng}, \&
  {Juric}}]{Kulkarni2014}
{Kulkarni}, S.~R., {Ofek}, E.~O., {Neill}, J.~D., {Zheng}, Z., \& {Juric}, M.
  2014, \apj, 797, 70

\bibitem[{{Law} {et~al.}(2017){Law}, {Abruzzo}, {Bassa}, {Bower},
  {Burke-Spolaor}, {Butler}, {Cantwell}, {Carey}, {Chatterjee}, {Cordes},
  {Demorest}, {Dowell}, {Fender}, {Gourdji}, {Grainge}, {Hessels}, {Hickish},
  {Kaspi}, {Lazio}, {McLaughlin}, {Michilli}, {Mooley}, {Perrott}, {Ransom},
  {Razavi-Ghods}, {Rupen}, {Scaife}, {Scott}, {Scholz}, {Seymour}, {Spitler},
  {Stovall}, {Tendulkar}, {Titterington}, {Wharton}, \& {Williams}}]{Law2017}
{Law}, C.~J., {Abruzzo}, M.~W., {Bassa}, C.~G., {et~al.} 2017, \apj, 850, 76

\bibitem[{{Li} {et~al.}(2020){Li}, {Lin}, {Xiong}, {Ge}, {Li}, {Li}, {Lu},
  {Zhang}, {Tuo}, {Nang}, {Zhang}, {Xiao}, {Chen}, {Song}, {Xu}, {Liu}, {Jia},
  {Cao}, {Zhang}, {Qu}, {Liao}, {Zhao}, {Tan}, {Nie}, {Zhao}, {Zheng}, {Zheng},
  {Luo}, {Cai}, {Li}, {Xue}, {Bu}, {Chang}, {Chen}, {Chen}, {Chen}, {Chen},
  {Chen}, {Cui}, {Cui}, {Deng}, {Dong}, {Du}, {Fu}, {Gao}, {Gao}, {Gao}, {Gu},
  {Guan}, {Guo}, {Han}, {Huang}, {Huo}, {Jiang}, {Jiang}, {Jin}, {Jin}, {Kong},
  {Li}, {Li}, {Li}, {Li}, {Li}, {Li}, {Li}, {Liang}, {Liu}, {Liu}, {Liu},
  {Liu}, {Liu}, {Lu}, {Lu}, {Luo}, {Ma}, {Meng}, {Ou}, {Sai}, {Shang}, {Song},
  {Sun}, {Tao}, {Wang}, {Wang}, {Wang}, {Wang}, {Wang}, {Wen}, {Wu}, {Wu},
  {Wu}, {Xiao}, {Yang}, {Yang}, {Yang}, {Yang}, {Yi}, {Yin}, {You}, {Zhang},
  {Zhang}, {Zhang}, {Zhang}, {Zhang}, {Zhang}, {Zhang}, {Zhang}, {Zhang},
  {Zhang}, {Zhang}, {Zhang}, {Zhang}, {Zhang}, {Zhang}, {Zhang}, {Zhou},
  {Zhou}, {Zhu}, {Zhu}, \& {Zhuang}}]{Li2020}
{Li}, C.~K., {Lin}, L., {Xiong}, S.~L., {et~al.} 2020, arXiv e-prints,
  arXiv:2005.11071

\bibitem[{{Li} {et~al.}(2019){Li}, {Zhang}, {Nagamine}, \& {Shi}}]{Li2019}
{Li}, Y., {Zhang}, B., {Nagamine}, K., \& {Shi}, J. 2019, \apjl, 884, L26

\bibitem[{{Lien} {et~al.}(2014){Lien}, {Barthelmy}, {Baumgartner}, {Cummings},
  {Gehrels}, {Krimm}, {Markwardt}, {Palmer}, {Sakamoto}, {Stamatikos},
  {Tueller}, \& {Ukwatta}}]{Lien2014}
{Lien}, A.~Y., {Barthelmy}, S.~D., {Baumgartner}, W.~H., {et~al.} 2014, GRB
  Coordinates Network, 16522, 1

\bibitem[{{Lin} {et~al.}(2020{\natexlab{a}}){Lin},
  {G{\"o}{\u{g}}{\"u}{\textcommabelow s}}, {Roberts}, {Kouveliotou}, {Kaneko},
  {van der Horst}, \& {Younes}}]{Lin2020a}
{Lin}, L., {G{\"o}{\u{g}}{\"u}{\textcommabelow s}}, E., {Roberts}, O.~J.,
  {et~al.} 2020{\natexlab{a}}, \apj, 893, 156

\bibitem[{{Lin} {et~al.}(2020{\natexlab{b}}){Lin}, {Zhang}, {Wang}, {Gao},
  {Guan}, {Han}, {Jiang}, {Jiang}, {Lee}, {Li}, {Men}, {Miao}, {Niu}, {Niu},
  {Sun}, {Wang}, {Wang}, {Xu}, {Xu}, {Xu}, {Yang}, {Yang}, {Yu}, {Zhang},
  {Zhang}, {Zhou}, {Zhu}, {Castro-Tirado}, {Dai}, {Ge}, {Hu}, {Li}, {Li}, {Li},
  {Liang}, {Jia}, {Querel}, {Shao}, {Wang}, {Wang}, {Wu}, {Xiong}, {Xu},
  {Yang}, {Zhang}, {Zhang}, {Zheng}, \& {Zou}}]{Lin2020}
{Lin}, L., {Zhang}, C.~F., {Wang}, P., {et~al.} 2020{\natexlab{b}}, arXiv
  e-prints, arXiv:2005.11479

\bibitem[{Lorimer {et~al.}(2007)Lorimer, Bailes, McLaughlin, Narkevic, \&
  Crawford}]{Lorimer2007}
Lorimer, D.~R., Bailes, M., McLaughlin, M.~A., Narkevic, D.~J., \& Crawford, F.
  2007, Science, 318, 777

\bibitem[{{Lu} \& {Kumar}(2018)}]{Lu2018}
{Lu}, W., \& {Kumar}, P. 2018, \mnras, 477, 2470

\bibitem[{{Lu} {et~al.}(2020){Lu}, {Kumar}, \& {Zhang}}]{Lu2020}
{Lu}, W., {Kumar}, P., \& {Zhang}, B. 2020, arXiv e-prints, arXiv:2005.06736

\bibitem[{{Lyubarsky}(2014)}]{Lyubarsky2014}
{Lyubarsky}, Y. 2014, \mnras, 442, L9

\bibitem[{{Lyutikov} \& {Popov}(2020)}]{Lyutikov2020}
{Lyutikov}, M., \& {Popov}, S. 2020, arXiv e-prints, arXiv:2005.05093

\bibitem[{{Marcote} {et~al.}(2017){Marcote}, {Paragi}, {Hessels}, {Keimpema},
  {van Langevelde}, {Huang}, {Bassa}, {Bogdanov}, {Bower}, {Burke-Spolaor},
  {Butler}, {Campbell}, {Chatterjee}, {Cordes}, {Demorest}, {Garrett}, {Ghosh},
  {Kaspi}, {Law}, {Lazio}, {McLaughlin}, {Ransom}, {Salter}, {Scholz},
  {Seymour}, {Siemion}, {Spitler}, {Tendulkar}, \& {Wharton}}]{Marcote2017}
{Marcote}, B., {Paragi}, Z., {Hessels}, J.~W.~T., {et~al.} 2017, \apjl, 834, L8

\bibitem[{{Marcote} {et~al.}(2020){Marcote}, {Nimmo}, {Hessels}, {Tendulkar},
  {Bassa}, {Paragi}, {Keimpema}, {Bhardwaj}, {Karuppusamy}, {Kaspi}, {Law},
  {Michilli}, {Aggarwal}, {Andersen}, {Archibald}, {Bandura}, {Bower}, {Boyle},
  {Brar}, {Burke-Spolaor}, {Butler}, {Cassanelli}, {Chawla}, {Demorest},
  {Dobbs}, {Fonseca}, {Giri}, {Good}, {Gourdji}, {Josephy}, {Kirichenko},
  {Kirsten}, {Landecker}, {Lang}, {Lazio}, {Li}, {Lin}, {Linford}, {Masui},
  {Mena-Parra}, {Naidu}, {Ng}, {Patel}, {Pen}, {Pleunis}, {Rafiei-Ravandi},
  {Rahman}, {Renard}, {Scholz}, {Siegel}, {Smith}, {Stairs}, {Vanderlinde}, \&
  {Zwaniga}}]{Marcote2020}
{Marcote}, B., {Nimmo}, K., {Hessels}, J.~W.~T., {et~al.} 2020, \nat, 577, 190

\bibitem[{{Margalit} {et~al.}(2020{\natexlab{a}}){Margalit}, {Beniamini},
  {Sridhar}, \& {Metzger}}]{Margalit2020}
{Margalit}, B., {Beniamini}, P., {Sridhar}, N., \& {Metzger}, B.~D.
  2020{\natexlab{a}}, arXiv e-prints, arXiv:2005.05283

\bibitem[{{Margalit} \& {Metzger}(2018)}]{Margalit2018}
{Margalit}, B., \& {Metzger}, B.~D. 2018, \apjl, 868, L4

\bibitem[{{Margalit} {et~al.}(2020{\natexlab{b}}){Margalit}, {Metzger}, \&
  {Sironi}}]{Margalit2020a}
{Margalit}, B., {Metzger}, B.~D., \& {Sironi}, L. 2020{\natexlab{b}}, \mnras,
  494, 4627

\bibitem[{{Mereghetti} {et~al.}(2020){Mereghetti}, {Savchenko}, {Ferrigno},
  {G{\"o}tz}, {Rigoselli}, {Tiengo}, {Bazzano}, {Bozzo}, {Coleiro},
  {Courvoisier}, {Doyle}, {Goldwurm}, {Hanlon}, {Jourdain}, {von Kienlin},
  {Lutovinov}, {Martin-Carrillo}, {Molkov}, {Natalucci}, {Onori}, {Panessa},
  {Rodi}, {Rodriguez}, {S{\'a}nchez-Fern{\'a}ndez}, {Sunyaev}, \&
  {Ubertini}}]{Mereghetti2020}
{Mereghetti}, S., {Savchenko}, V., {Ferrigno}, C., {et~al.} 2020, arXiv
  e-prints, arXiv:2005.06335

\bibitem[{{Metzger} {et~al.}(2017){Metzger}, {Berger}, \&
  {Margalit}}]{Metzger2017}
{Metzger}, B.~D., {Berger}, E., \& {Margalit}, B. 2017, \apj, 841, 14

\bibitem[{{Metzger} {et~al.}(2019){Metzger}, {Margalit}, \&
  {Sironi}}]{Metzger2019}
{Metzger}, B.~D., {Margalit}, B., \& {Sironi}, L. 2019, \mnras, 485, 4091

\bibitem[{{Michilli} {et~al.}(2018){Michilli}, {Seymour}, {Hessels}, {Spitler},
  {Gajjar}, {Archibald}, {Bower}, {Chatterjee}, {Cordes}, {Gourdji}, {Heald},
  {Kaspi}, {Law}, {Sobey}, {Adams}, {Bassa}, {Bogdanov}, {Brinkman},
  {Demorest}, {Fernand ez}, {Hellbourg}, {Lazio}, {Lynch}, {Maddox}, {Marcote},
  {McLaughlin}, {Paragi}, {Ransom}, {Scholz}, {Siemion}, {Tendulkar}, {van
  Rooy}, {Wharton}, \& {Whitlow}}]{Michilli2018}
{Michilli}, D., {Seymour}, A., {Hessels}, J.~W.~T., {et~al.} 2018, \nat, 553,
  182

\bibitem[{{Murase} {et~al.}(2016){Murase}, {Kashiyama}, \&
  {M{\'e}sz{\'a}ros}}]{Murase2016}
{Murase}, K., {Kashiyama}, K., \& {M{\'e}sz{\'a}ros}, P. 2016, \mnras, 461,
  1498

\bibitem[{{Oppermann} {et~al.}(2018){Oppermann}, {Yu}, \&
  {Pen}}]{Oppermann2018}
{Oppermann}, N., {Yu}, H.-R., \& {Pen}, U.-L. 2018, \mnras, 475, 5109

\bibitem[{{Ordog} {et~al.}(2019){Ordog}, {Booth}, {Van Eck}, {Brown}, \&
  {Landecker}}]{Ordog2019}
{Ordog}, A., {Booth}, R., {Van Eck}, C., {Brown}, J.-A., \& {Landecker}, T.
  2019, Galaxies, 7, 43

\bibitem[{{Petroff} {et~al.}(2019){Petroff}, {Hessels}, \&
  {Lorimer}}]{Petroff2019}
{Petroff}, E., {Hessels}, J.~W.~T., \& {Lorimer}, D.~R. 2019, \aapr, 27, 4

\bibitem[{{Platts} {et~al.}(2019){Platts}, {Weltman}, {Walters}, {Tendulkar},
  {Gordin}, \& {Kandhai}}]{Platts2019}
{Platts}, E., {Weltman}, A., {Walters}, A., {et~al.} 2019, \physrep, 821, 1

\bibitem[{{Popov} \& {Postnov}(2013)}]{Popov2013}
{Popov}, S.~B., \& {Postnov}, K.~A. 2013, arXiv e-prints, arXiv:1307.4924

\bibitem[{{Rajwade} {et~al.}(2020){Rajwade}, {Mickaliger}, {Stappers},
  {Morello}, {Agarwal}, {Bassa}, {Breton}, {Caleb}, {Karastergiou}, {Keane}, \&
  {Lorimer}}]{Rajwade2020}
{Rajwade}, K.~M., {Mickaliger}, M.~B., {Stappers}, B.~W., {et~al.} 2020,
  \mnras, arXiv:2003.03596

\bibitem[{{Reid} \& {Ratcliffe}(2014)}]{Reid2014}
{Reid}, H. A.~S., \& {Ratcliffe}, H. 2014, Research in Astronomy and
  Astrophysics, 14, 773

\bibitem[{{Ridnaia} {et~al.}(2020){Ridnaia}, {Svinkin}, {Frederiks}, {Bykov},
  {Popov}, {Aptekar}, {Golenetskii}, {Lysenko}, {Tsvetkova}, {Ulanov}, \&
  {Cline}}]{Ridnaia2020}
{Ridnaia}, A., {Svinkin}, D., {Frederiks}, D., {et~al.} 2020, arXiv e-prints,
  arXiv:2005.11178

\bibitem[{{Stamatikos} {et~al.}(2014){Stamatikos}, {Malesani}, {Page}, \&
  {Sakamoto}}]{Stamatikos2014}
{Stamatikos}, M., {Malesani}, D., {Page}, K.~L., \& {Sakamoto}, T. 2014, GRB
  Coordinates Network, 16520, 1

\bibitem[{{Tavani} {et~al.}(2020){Tavani}, {Casentini}, {Ursi}, {Verrecchia},
  {Addis}, {Antonelli}, {Argan}, {Barbiellini}, {Baroncelli}, {Bernardi},
  {Bianchi}, {Bulgarelli}, {Caraveo}, {Cardillo}, {Cattaneo}, {Chen}, {Costa},
  {Del Monte}, {Di Cocco}, {Di Persio}, {Donnarumma}, {Evangelista}, {Feroci},
  {Ferrari}, {Fioretti}, {Fuschino}, {Galli}, {Gianotti}, {Giuliani},
  {Labanti}, {Lazzarotto}, {Lipari}, {Longo}, {Lucarelli}, {Magro},
  {Marisaldi}, {Mereghetti}, {Morelli}, {Morselli}, {Naldi}, {Pacciani},
  {Parmiggiani}, {Paoletti}, {Pellizzoni}, {Perri}, {Perotti}, {Piano},
  {Picozza}, {Pilia}, {Pittori}, {Puccetti}, {Pupillo}, {Rapisarda},
  {Rappoldi}, {Rubini}, {Setti}, {Soffitta}, {Trifoglio}, {Trois},
  {Vercellone}, {Vittorini}, {Giommi}, \& {D' Amico}}]{Tavani2020}
{Tavani}, M., {Casentini}, C., {Ursi}, A., {et~al.} 2020, arXiv e-prints,
  arXiv:2005.12164

\bibitem[{{The CHIME/FRB Collaboration} {et~al.}(2020){The CHIME/FRB
  Collaboration}, {:}, {Andersen}, {Band ura}, {Bhardwaj}, {Bij}, {Boyce},
  {Boyle}, {Brar}, {Cassanelli}, {Chawla}, {Chen}, {Cliche}, {Cook},
  {Cubranic}, {Curtin}, {Denman}, {Dobbs}, {Dong}, {Fandino}, {Fonseca},
  {Gaensler}, {Giri}, {Good}, {Halpern}, {Hill}, {Hinshaw}, {H{\"o}fer},
  {Josephy}, {Kania}, {Kaspi}, {Landecker}, {Leung}, {Li}, {Lin}, {Masui},
  {Mckinven}, {Mena-Parra}, {Merryfield}, {Meyers}, {Michilli}, {Milutinovic},
  {Mirhosseini}, {M{\"u}nchmeyer}, {Naidu}, {Newburgh}, {Ng}, {Patel}, {Pen},
  {Pinsonneault-Marotte}, {Pleunis}, {Quine}, {Rafiei-Ravandi}, {Rahman},
  {Ransom}, {Renard}, {Sanghavi}, {Scholz}, {Shaw}, {Shin}, {Siegel}, {Singh},
  {Smegal}, {Smith}, {Stairs}, {Tan}, {Tendulkar}, {Tretyakov}, {Vanderlinde},
  {Wang}, {Wulf}, \& {Zwaniga}}]{CHIME/FRB2020}
{The CHIME/FRB Collaboration}, {:}, {Andersen}, B.~C., {et~al.} 2020, arXiv
  e-prints, arXiv:2005.10324

\bibitem[{{Thornton} {et~al.}(2013){Thornton}, {Stappers}, {Bailes},
  {Barsdell}, {Bates}, {Bhat}, {Burgay}, {Burke-Spolaor}, {Champion}, {Coster},
  {D'Amico}, {Jameson}, {Johnston}, {Keith}, {Kramer}, {Levin}, {Milia}, {Ng},
  {Possenti}, \& {van Straten}}]{Thornton2013}
{Thornton}, D., {Stappers}, B., {Bailes}, M., {et~al.} 2013, Science, 341, 53

\bibitem[{{Veres} {et~al.}(2020){Veres}, {Bissaldi}, {Briggs}, \& {Fermi GBM
  Team}}]{Veres2020}
{Veres}, P., {Bissaldi}, E., {Briggs}, M.~S., \& {Fermi GBM Team}. 2020, GRB
  Coordinates Network, 27531, 1

\bibitem[{{Wadiasingh} \& {Timokhin}(2019)}]{Wadiasingh2019}
{Wadiasingh}, Z., \& {Timokhin}, A. 2019, \apj, 879, 4

\bibitem[{{Wang} {et~al.}(2020){Wang}, {Wang}, {Yang}, {Yu}, {Zuo}, \&
  {Dai}}]{Wang2020}
{Wang}, F.~Y., {Wang}, Y.~Y., {Yang}, Y.-P., {et~al.} 2020, \apj, 891, 72

\bibitem[{{Wang} \& {Yu}(2017)}]{Wang2017}
{Wang}, F.~Y., \& {Yu}, H. 2017, \jcap, 03, 023

\bibitem[{{Wang}(2020)}]{WangJ2020}
{Wang}, J.-S. 2020, arXiv e-prints, arXiv:2006.14503

\bibitem[{{Yang} \& {Zhang}(2018)}]{Yang2018}
{Yang}, Y.-P., \& {Zhang}, B. 2018, \apj, 868, 31

\bibitem[{{Yang} {et~al.}(2020){Yang}, {Zhu}, {Zhang}, \& {Wu}}]{Yang2020}
{Yang}, Y.-P., {Zhu}, J.-P., {Zhang}, B., \& {Wu}, X.-F. 2020, arXiv e-prints,
  arXiv:2006.03270

\bibitem[{{Younes} {et~al.}(2017){Younes}, {Kouveliotou}, {Jaodand}, {Baring},
  {van der Horst}, {Harding}, {Hessels}, {Gehrels}, {Gill}, {Huppenkothen},
  {Granot}, {G{\"o}{\u{g}}{\"u}{\textcommabelow s}}, \& {Lin}}]{Younes2017}
{Younes}, G., {Kouveliotou}, C., {Jaodand}, A., {et~al.} 2017, \apj, 847, 85

\bibitem[{{Yu} {et~al.}(2020){Yu}, {Zou}, {Dai}, \& {Yu}}]{Yu2020}
{Yu}, Y.-W., {Zou}, Y.-C., {Dai}, Z.-G., \& {Yu}, W.-F. 2020, arXiv e-prints,
  arXiv:2006.00484

\bibitem[{{Zhang} {et~al.}(2020){Zhang}, {Jiang}, {Men}, {Wang}, {Xu}, {Xu},
  {Niu}, {Zhou}, {Guan}, {Han}, {Jiang}, {Lee}, {Li}, {Lin}, {Niu}, {Wang},
  {Wang}, {Xu}, {Yu}, {Zhang}, \& {Zhu}}]{FAST2020}
{Zhang}, C.~F., {Jiang}, J.~C., {Men}, Y.~P., {et~al.} 2020, The Astronomer's
  Telegram, 13699, 1

\bibitem[{{Zhang} {et~al.}(2019){Zhang}, {Wang}, \& {Dai}}]{ZhangG2019}
{Zhang}, G.~Q., {Wang}, F.~Y., \& {Dai}, Z.~G. 2019, arXiv e-prints,
  arXiv:1903.11895

\bibitem[{{Zhang} {et~al.}(2018){Zhang}, {Gajjar}, {Foster}, {Siemion},
  {Cordes}, {Law}, \& {Wang}}]{ZhangY2018}
{Zhang}, Y.~G., {Gajjar}, V., {Foster}, G., {et~al.} 2018, \apj, 866, 149

\bibitem[{{Zhong} {et~al.}(2020){Zhong}, {Dai}, {Zhang}, \& {Deng}}]{Zhong2020}
{Zhong}, S.-Q., {Dai}, Z.-G., {Zhang}, H.-M., \& {Deng}, C.-M. 2020, \apjl,
  898, L5

\bibitem[{{Zhou} {et~al.}(2020){Zhou}, {Zhou}, {Chen}, {Wang}, {Vink}, \&
  {Wang}}]{Zhou2020}
{Zhou}, P., {Zhou}, X., {Chen}, Y., {et~al.} 2020, arXiv e-prints,
  arXiv:2005.03517

\end{thebibliography}

\end{document}